# Learning without Recall in Directed Circles and Rooted Trees

Mohammad Amin Rahimian, Ali Jadbabaie *

*Abstract*—This work investigates the case of a network of agents that attempt to learn some unknown state of the world amongst the finitely many possibilities. At each time step, agents all receive random, independently distributed private signals whose distributions are dependent on the unknown state of the world. However, it may be the case that some or any of the agents cannot distinguish between two or more of the possible states based only on their private observations, as when several states result in the same distribution of the private signals. In our model, the agents form some initial belief (probability distribution) about the unknown state and then refine their beliefs in accordance with their private observations, as well as the beliefs of their neighbors. An agent learns the unknown state when her belief converges to a point mass that is concentrated at the true state. A rational agent would use the Bayes' rule to incorporate her neighbors' beliefs and own private signals over time. While such repeated applications of the Bayes' rule in networks can become computationally intractable; in this paper, we show that in the canonical cases of directed star, circle or path networks and their combinations, one can derive a class of memoryless update rules that replicate that of a single Bayesian agent but replace the self beliefs with the beliefs of the neighbors. This way, one can realize an exponentially fast rate of learning similar to the case of Bayesian (fully rational) agents. The proposed rules are a special case of the *Learning without Recall* approach that we develop in a companion paper, and it has the advantage that while preserving essential features of the Bayesian inference, they are made tractable. In particular, the agents can rely on the observational abilities of their neighbors and their neighbors' neighbors etc. to learn the unknown state; even though they themselves cannot distinguish the truth.

## I. Introduction & Background

Consider a group of agents who try to estimate an unknown state of the world. Each agent receives a sequence of independent and identically distributed (i.i.d.) private signals whose distribution is determined by the unknown state. Suppose further that the belief of each agent about the unknown state is represented by a discrete probability distribution over the finitely many possibilities, and that every agent sequentially applies the Bayes' rule to her observations at each step, and updates her beliefs accordingly. It is a well-known consequence of the classical results in merging and learning theory [1], [2] that the beliefs formed in the above manner constitute a bounded martingale and converge to a limiting distribution as the number of observations increases. However, the limiting distribution may differ from a point mass centered at the truth, in which case the agent fails to learn the true state asymptotically. This may be the case, for instance if the agent faces an identification problem, that is when there are states other than the true state which are observationally equivalent to the true state and induce the same distribution on her sequence of privately observed signals. Accordingly, the agents have an incentive to communicate in a social network so that they can resolve their identification problems by relying on each other's observational abilities. This leads to the problem of social learning that is a classical focus of behavioral microeconomic theory [3], [4] and has close parallels in distributed estimation and statistical learning theory [5], [6], [7], [8], [9].

Rational agents in a social network would apply the Bayes' rule successively to their observations at each step, which include not only their private signals but also the beliefs communicated by their neighbors. However, such repeated applications of Bayes' rule in networks become computationally intractable especially if the agents are unaware of the global network structure. This is due to the fact that the agents at each step should use their local data that is increasing with time, and make inferences about the global signal structures that can have led to their observations. Indeed, tractable modeling and analysis of rational behavior in networks is an important problem in Bayesian economics and have attracted much attention [10], [11]. On the other side of the spectrum are the literature such as [12], [13] which attempt to investigate the problem of learning in networks through iterative applications of some update rules that do not necessarily result in the Bayesian beliefs, but can nonetheless provide the asymptotic properties of learning and consensus under certain conditions.

In this work, we first consider the behavior of a single Bayesian agent that observes a sequence of i.i.d. signals conditioned on the unknown state of the world. We show that the learning rate for such an agent is exponentially fast with an asymptotic rate that can be expressed in terms of the relative entropies between the likelihood structures of her signals under various states of the world. We next use these results to upper bound the rate of learning for a Bayesian agent in a social network observing not only her private signals but also her neighbor's beliefs. The focus is then restricted to the case of directed circles, for which we propose a class of update rules offering exponentially fast learning at an asymptotic rate that is within a constant factor $1/l$ of the derived upper bound, $l$ being the length of the circle. These updates are also applied to other hybrid structures, where the center node of rooted tree is replaced by a circle. These updates are a special case of the *Learning without Recall* rules, which we develop in a companion paper.

The remainder of this paper is organized as follows. The modeling and formulation are set forth in Section II. The case of a single Bayesian agent is investigated in Section III. Learning without Recall updates for directed circle and rooted trees together with their asymptotic properties including the learning and convergence rate are then presented in Section IV, and the paper is concluded by Section V.

* The authors are with the Department of Electrical and Systems Engineering, University of Pennsylvania, Philadelphia, PA 19104-6228 USA (email: jadbabai@seas.upenn.edu). This work was supported by ARO MURI W911NF-12-1-0509.

## II. THE MODEL

*Notation:* Throughout the paper, $\mathbb{R}$ is the set of real numbers, $\mathbb{N}$ denotes the set of all natural numbers, and $\mathbb{W} = \mathbb{N} \cup \{0\}$. For $n \in \mathbb{N}$ a fixed integer the set of integers $\{1, 2, \ldots, n\}$ is denoted by $[n]$, while any other set is represented by a calligraphic capital letter. The cardinality of a set $\mathcal{X}$, which is the number of its elements, is denoted by $|\mathcal{X}|$, and $\mathscr{P}(\mathcal{X}) = \{\mathcal{M}; \mathcal{M} \subset \mathcal{X}\}$ denotes the power-set of $\mathcal{X}$, which is the set of all its subsets. The difference of two sets $\mathcal{X}$ and $\mathcal{Y}$ is defined by $\mathcal{X} \backslash \mathcal{Y} := \{x; x \in \mathcal{X} \text{ and } x \notin \mathcal{Y}\}$. Boldface letters denote random variables.

Consider a set of $n$ agents that are labeled by $[n]$ and interact according to a directed information flow structure given by a digraph $\mathcal{G} = ([n], \mathcal{E})$, where $\mathcal{E} \subset [n] \times [n]$ is the set of directed edges. $\mathcal{N}(i) = \{j \in [n]; (j, i) \in \mathcal{E}\}$ is called the neighborhood of agent $i$ and is the set of all agents whose beliefs are observed by agent $i$, and $\deg(i) = |\mathcal{N}(i)|$ is called the degree of agent $i$. For $l, m \in [n]$, a path $\mathcal{P}_k(l, m)$ of length $k$ from $l$ to $m$ is a sequence of $k$ distinct integers $i_1, i_2, \ldots, i_k$, such that $i_1 = l$, $i_k = m$ and $(i_{j-1}, i_j) \in \mathcal{E}$ for all $j \in [k]$.

The set of finitely many possible states of the world is denoted by $\Theta$, and $\Delta \Theta$ is the space of all probability measures on the set $\Theta$. The goal is to decide amongst the finitely many possibilities in the state space $\Theta$. A random variable $\boldsymbol{\theta}$ is chosen randomly from $\Theta$ by the nature and according to the probability measure $\nu(\cdot) \in \Delta \Theta$, which satisfies $\nu(\hat{\theta}) > 0, \forall \hat{\theta} \in \Theta$ and is referred to as the common prior. Associated with each agent $i$, $\mathcal{S}_i$ is a finite set called the signal space of $i$, and given $\boldsymbol{\theta}$, $\ell_i(\cdot \mid \boldsymbol{\theta})$ is a probability measure on $\mathcal{S}_i$, which is referred to as the *signal structure* or *likelihood function* of agent $i$. Furthermore, $(\Omega, \mathscr{F}, \mathbb{P})$ is a probability triplet, where

$$\Omega = \Theta \times \left(\prod_{i \in [n]} \mathcal{S}_i\right)^{\mathbb{W}},$$

is an infinite product space with a general element $\omega = (\theta; (s_{1,0}, \ldots, s_{n,0}), (s_{1,1}, \ldots, s_{n,1}), \ldots)$ and the associated sigma field $\mathscr{F} = \mathscr{P}(\Omega)$. $\mathbb{P}(\cdot)$ is the probability measure on $\Omega$ which assigns probabilities consistently with the common prior $\nu(\cdot)$ and the likelihood functions $\ell_i(\cdot \mid \boldsymbol{\theta}), i \in [n]$, and in such a way that conditional on $\boldsymbol{\theta}$ the random variables $\{\mathbf{s}_{i,t}, t \in \mathbb{W}, i \in [n]\}$ are independent. $\mathbb{E}\{\cdot\}$ is the expectation operator, which represents integration with respect to $d\mathbb{P}(\omega)$.

### A. Signals

Let $t \in \mathbb{W}$ denote the time index and for each agent $i$, define $\{\mathbf{s}_{i,t}, t \in \mathbb{W}\}$ to be a sequence of independent and identically distributed random variables with the probability mass function $\ell_i(\cdot \mid \boldsymbol{\theta})$; this sequence represents the private observations made by agent $i$ at each time period $t$. The privately observed signals are independent and identically distributed across time; and at any given time $t$, each agent makes a signal observation that is independent of the rest of the agents. In particular, for $t \in \mathbb{W}$ and $\hat{\theta} \in \Theta$ both fixed, let $\mathcal{L}(\cdot \mid \hat{\theta})$ denote the joint law for the random vector $(\mathbf{s}_{1,t}, \ldots, \mathbf{s}_{n,t})$, then $\mathcal{L}(\cdot \mid \hat{\theta})$ does not depend on $t$, and it factors

$$\mathcal{L}(\cdot \mid \hat{\theta}) = \prod_{i \in [n]} \ell_i(\cdot \mid \hat{\theta}), \quad (1)$$

as the product probability measure on the product space $\prod_{i \in [n]} \mathcal{S}_i$. The signal structures and the joint law given in (1), as well as the common prior $\nu(\cdot)$ and their corresponding sample spaces $\prod_{i \in [n]} \mathcal{S}_i$ and $\Theta$ are common knowledge amongst all the agents. The assumption of common knowledge in the case of fully rational (Bayesian) agents implies that given the same observations of one another's beliefs or private signals distinct agents would make identical inferences; in the sense that starting form the same belief about the unknown $\boldsymbol{\theta}$, their updated beliefs given the same observations, would be the same.

### B. Beliefs

For each time instant $t$, let $\boldsymbol{\mu}_{i,t}(\cdot)$ be the probability mass function on $\Theta$, representing the *opinion* or *belief* at time $t$ of agent $i$ about the realized value of $\boldsymbol{\theta}$. Note that $\boldsymbol{\mu}_{i,t}(\cdot)$ is random since it depends on the random observations of the agent. The goal is to investigate the problem of asymptotic learning, i.e. for each agents to learn the true realized value $\theta \in \Theta$ of $\boldsymbol{\theta}$ asymptotically. That is to have $\boldsymbol{\mu}_{i,t}(\cdot)$ to converge to a point mass centered at $\theta$, where the convergence could be in probability or in the $\mathbb{P}$-almost sure sense.

At $t = 0$ the values $\boldsymbol{\theta} = \theta$, followed by $s_i \in \mathcal{S}_i$ of $\mathbf{s}_{i,0}$ are realized and the latter is observed by agent $i$ for all $i \in [n]$, who then forms an initial Bayesian opinion $\boldsymbol{\mu}_{i,0}(\cdot)$ about the value of $\theta$. Given $\mathbf{s}_{i,0}$, and using the Bayes' rule for each agent $i \in [n]$, the initial belief in terms of the observed signal $\mathbf{s}_{i,0}$ is given by:

$$\boldsymbol{\mu}_{i,0}(\hat{\theta}) = \frac{\nu(\hat{\theta}) \ell_i(\mathbf{s}_{i,0} \mid \hat{\theta})}{\sum_{\tilde{\theta} \in \Theta} \nu(\tilde{\theta}) \ell_i(\mathbf{s}_{i,0} \mid \tilde{\theta})}. \quad (2)$$

At any successive time step $t \geqslant 1$, each agent $i$ observes the realized values of $\mathbf{s}_{i,t}$ as well as the current beliefs of its neighbors $\boldsymbol{\mu}_{k,t-1}(\cdot), \forall k \in \mathcal{N}(i)$ and forms a refined opinion $\boldsymbol{\mu}_{i,t}(\cdot)$ by incorporating all the data that have been made available to her by the time $t$.

## III. THE CASE OF A SINGLE BAYESIAN AGENT

A Bayesian agent $i$ that starts with a prior $\nu(\cdot)$ on the state of the world and successively uses the Bayes rule to update her beliefs based on the signals $\{\mathbf{s}_{i,t}, t \in \mathbb{N}\}$ that she observes would form the initial belief given in (2) and will then sequentially update her beliefs according to the Bayes' rule:

$$\boldsymbol{\mu}_{i,t}(\hat{\theta}) = \frac{\boldsymbol{\mu}_{i,t-1}(\hat{\theta}) \ell_i(\mathbf{s}_{i,t} \mid \hat{\theta})}{\sum_{\tilde{\theta} \in \Theta} \boldsymbol{\mu}_{i,t-1}(\tilde{\theta}) \ell_i(\mathbf{s}_{i,t} \mid \tilde{\theta})}, \forall \hat{\theta} \in \Theta. \quad (3)$$

For any false state $\check{\theta} \in \Theta \backslash \{\theta\}$ let $\mathbf{r}_{i,t}(\check{\theta}) := \log\left(\ell_i(\mathbf{s}_{i,t} \mid \check{\theta})/\ell_i(\mathbf{s}_{i,t} \mid \theta)\right)$, be the random variable representing the log-likelihood ratio of the private signal that agent $i$ observes at time $t$ and under the false state $\check{\theta}$; and similarly, let $\boldsymbol{\lambda}_{i,t}(\check{\theta}) := \log\left(\boldsymbol{\mu}_{i,t}(\check{\theta})/\boldsymbol{\mu}_{i,t}(\theta)\right)$ be the log-likelihood ratio of the belief of agent $i$ at time $t$ under the false state $\check{\theta}$. Note that $\boldsymbol{\lambda}_{i,t}(\check{\theta})$ and $\boldsymbol{\mu}_{i,t}(\check{\theta})$ are related through

$$\boldsymbol{\mu}_{i,t}(\check{\theta}) = \frac{e^{\boldsymbol{\lambda}_{i,t}(\check{\theta})}}{1 + \sum_{\tilde{\theta} \in \Theta \setminus \{\theta\}} e^{\boldsymbol{\lambda}_{i,t}(\tilde{\theta})}}, \quad (4)$$

and the Bayesian belief update in (3) translates into the following linear update for the log-likelihood ratios: $\boldsymbol{\lambda}_{i,t}(\check{\theta}) = \boldsymbol{\lambda}_{i,t-1}(\check{\theta}) + \mathbf{r}_{i,t}(\check{\theta})$, which leads to

$$\boldsymbol{\lambda}_{i,t}(\check{\theta}) = \boldsymbol{\lambda}_{i,0}(\check{\theta}) + \sum_{q=1}^{n} \mathbf{r}_{i,q}(\check{\theta}). \quad (5)$$

Next note that for all $t \in \mathbb{N}$,

$$\mathbb{E}\left\{\mathbf{r}_{i,t}(\check{\theta})\right\} = \sum_{s_i \in \mathcal{S}_i} \ell_i(s_i|\theta) \log\left(\frac{\ell_i(s_i|\check{\theta})}{\ell_i(s_i|\theta)}\right)$$
$$:= -D_{KL}\left(\ell_i(\cdot|\theta) || \ell_i(\cdot|\check{\theta})\right) \leqslant 0,$$

where the inequality follows from the positivity of the Kullback-Leibler divergence $D_{KL}(\cdot||\cdot)$ and is strict whenever $\ell_i(\cdot|\check{\theta}) \not\equiv \ell_i(\cdot|\theta)$, i.e. $\exists s \in \mathcal{S}_i$ such that $\ell_i(s|\check{\theta}) \neq \ell_i(s|\theta)$ [14, Theorem 2.6.3]. In particular $\mathbf{r}_{i,t}(\check{\theta}), t \in \mathbb{N}$ are integrable, independent and identically distributed variables, thence by the Kolmogrov's strong law of large number we get

$$\frac{1}{t} \sum_{q=1}^{t} \mathbf{r}_{i,q}(\check{\theta}) \to \mathbb{E}\left\{\mathbf{r}_{i,t}(\check{\theta})\right\}, \quad (6)$$

$\mathbb{P}$-almost surely. This in turn implies that if $\mathbb{E}\left\{\mathbf{r}_{i,t}(\check{\theta})\right\} = -D_{KL}\left(\ell_i(\cdot|\theta)||\ell_i(\cdot|\check{\theta})\right) < 0$ or equivalently $\ell_i(\cdot|\check{\theta}) \not\equiv \ell_i(\cdot|\theta)$, then $\boldsymbol{\lambda}_{i,t}(\check{\theta}) \to -\infty$. Substituting the latter in (4) then yields that $\boldsymbol{\mu}_{i,t}(\check{\theta}) \to 0$, $\mathbb{P}$-almost surely; with probability one, the agent asymptotically rejects any false state $\check{\theta} \in \Theta \setminus \{\theta\}$ satisfying $D_{KL}\left(\ell_i(\cdot|\theta)||\ell_i(\cdot|\check{\theta})\right) > 0$. In particular we have,

**Theorem 1.** *If $\forall \check{\theta} \in \Theta \setminus \{\theta\}$, $D_{KL}\left(\ell_i(\cdot|\theta)||\ell_i(\cdot|\check{\theta})\right) > 0$, then $\boldsymbol{\mu}_{i,t}(\theta) \to 1$; $\mathbb{P}$-almost surely, under the update rule in (3) and the specified model.*

**Remark 1.** *It is instructive to regard the almost sure convergence of beliefs stated for a Bayesian agent in Theorem 1 as a consequence of the bounded martingale convergence theorem. To see how, for all $\omega \in \Omega$ and $\hat{\theta} \in \Theta$, let $\mathbb{1}_{\boldsymbol{\theta}=\hat{\theta}}(\omega)$ be the indicator variable for the true state of the world being $\hat{\theta}$, i.e. $\mathbb{1}_{\boldsymbol{\theta}=\hat{\theta}}(\omega) = 1$ if $\boldsymbol{\theta}(\omega) = \hat{\theta}$ and $\mathbb{1}_{\boldsymbol{\theta}=\hat{\theta}}(\omega) = 0$, otherwise. For each $t \in \mathbb{W}$, define $\mathscr{F}_{i,t} = \sigma\left(\mathbf{s}_{i,t}, \mathbf{s}_{i,t-t}, \mathbf{s}_{i,t-2}, \ldots, \mathbf{s}_{i,1}, \mathbf{s}_{i,0}\right)$ as the sigma fields generated by the private signals of the agent upto time $t$. Note in particular that $\mathscr{F}_{0,t} = \sigma(\{\mathbf{s}_{i,0}\})$ and $\{\mathscr{F}_{i,t}, t \in \mathbb{W}\}$ is a filtration on the measure space $(\Omega, \mathscr{F})$. The Bayes rule in (2) can be interpreted as*

$$\boldsymbol{\mu}_{i,0}(\hat{\theta}) = \mathbb{P}\{\boldsymbol{\theta} = \hat{\theta} \mid \mathbf{s}_{i,0}\} = \mathbb{E}\{\mathbb{1}_{\boldsymbol{\theta}=\hat{\theta}} \mid \mathscr{F}_{i,0}\}. \quad (7)$$

*Similarly, starting from the Bayesian opinion in (7), application of (3) at times $t > 0$ is exactly the Bayesian update of agent $i$'s belief from time $t-1$ to time $t$, given that at time $t$ agent $i$ has observed the signal $\mathbf{s}_{i,t}$. Whence, (3) can be combined with (7) to get $\boldsymbol{\mu}_{i,t}(\hat{\theta}) = \mathbb{E}\{\mathbb{1}_{\boldsymbol{\theta}=\hat{\theta}} \mid \mathscr{F}_{i,t}\}$, $\forall t \geqslant 0$. The beliefs form a bounded martingale with respect to the filtration introduced above, and it is exactly the setting for the martingale convergence theorem. Indeed, the convergence of $\boldsymbol{\mu}_{i,t}(\hat{\theta})$ to $\mathbb{1}_{\boldsymbol{\theta}=\hat{\theta}}$ is now immediate, since $D_{KL}\left(\ell_i(\cdot|\theta)||\ell_i(\cdot|\check{\theta})\right) > 0, \forall \check{\theta} \neq \theta$ implies that $\{\boldsymbol{\theta} = \hat{\theta}\} \in \mathscr{F}_{i,\infty}$ and by Levy's zero-one law [15], $\lim_{t \to \infty} \boldsymbol{\mu}_{i,t}(\hat{\theta}) = \mathbb{E}\{\mathbb{1}_{\boldsymbol{\theta}=\hat{\theta}} \mid \mathscr{F}_{i,\infty}\} = \mathbb{1}_{\boldsymbol{\theta}=\hat{\theta}}$, $\mathbb{P}$-almost surely.*

By the presumption of Theorem 1 we are lead to define for any $\hat{\theta} \in \Theta$ the set of those states $\tilde{\theta} \in \Theta \setminus \{\hat{\theta}\}$ that are observationally equivalent to $\hat{\theta}$ for agent $i$ and denote it by

$$\mathcal{O}_i(\hat{\theta}) = \left\{\tilde{\theta} \in \Theta \setminus \{\hat{\theta}\} : D_{KL}\left(\ell_i(\cdot|\hat{\theta}) || \ell_i(\cdot|\tilde{\theta})\right) = 0\right\}.$$

In the case where $\mathcal{O}_i(\hat{\theta}) \neq \varnothing$, the statement of Theorem 1 can be refined as follows.

**Corollary 1.** *Under the update rule in (3) and the specified model, it holds true with $\mathbb{P}$-probability 1 that as $t \to \infty$, $\boldsymbol{\mu}_{i,t}(\check{\theta}) \to 0$, $\forall \check{\theta} \notin \mathcal{O}_i(\hat{\theta})$ and*

$$\boldsymbol{\mu}_{i,t}(\hat{\theta}) \to \frac{\nu(\hat{\theta})}{\displaystyle\sum_{\tilde{\theta} \in \mathcal{O}_i(\theta)} \nu(\tilde{\theta})}, \forall \hat{\theta} \in \mathcal{O}_i(\theta).$$

*A. Exponentially Fast Learning*

We can push the preceding results further to prove an exponential rate of convergence for the beliefs. Indeed, by applying (5) and (6) for the beliefs on the false state given in (4), it follows that for any $0 < \gamma < D_{KL}\left(\ell_i(\cdot|\theta)||\ell_i(\cdot|\check{\theta})\right)$ we can write $\boldsymbol{\mu}_{i,t}(\check{\theta}) = e^{-\gamma t}\mathbf{z}_t$, where $\{\mathbf{z}_t, t \in \mathbb{N}\}$ is a process satisfying $\mathbf{z}_t \to 0$ with $\mathbb{P}$-probability one as $t \to \infty$. Instead, if we write $\boldsymbol{\mu}_{i,t}(\theta) = 1 - \sum_{\check{\theta} \in \Theta \setminus \{\theta\}} \boldsymbol{\mu}_{i,t}(\check{\theta})$, then it follows that the almost sure convergence stated in Theorem 1 occurs with an exponentially fast asymptotic rate of

$$R_i(\theta) := \min_{\check{\theta} \in \Theta \setminus \{\theta\}} D_{KL}\left(\ell_i(\cdot|\theta)||\ell_i(\cdot|\check{\theta})\right). \quad (8)$$

In effect, the states of the world are to be statistically distinguished by the observed signals $\mathbf{s}_{i,t}, t \in \mathbb{W}$. Different states $\tilde{\theta} \in \Theta$ are distinguished through their different likelihood functions $\ell_i(\cdot \mid \hat{\theta})$ and the more refined such differences are, the better the states are distinguished. The asymptotic rate derived in (8) is one measure of resolution for the likelihood structure, or indeed for the filtration introduced in Remark 1.

*B. Bayesian Learning in Networks*

The preceding discussion laid the case for the exponentially fast learning of a single agent that applies the Bayes rule successively to her observed signals and updates her belief about the true state of the world at each round. One might suggest to use the same framework in a network setting where the agents have access to their neighbors' beliefs at successive time steps, by considering the conditional probabilities given both the neighbors' beliefs and the private signals. However, repeated applications of the Bayes' rule in networks become computationally intractable, partly due to the fact that each agent needs to use her local data that is increasing over time and make inferences about the global network structure, which is unknown to her. The complexities associated with the Bayesian framework has limited its application to the simplest networks such as three-link ones [16]. Apart from the difficulties associated with the network structure, the increasing history of the observations that a fully Bayesian agent needs to take into account imposes foreboding computational burden [17]. Nonetheless, an upper-bound for the exponential rate of learning by a particular agent $i$ in a network of Bayesian agents can be obtained as follows. Consider an outside Bayesian agent $\hat{i}$ who shares the same common knowledge of the prior and signal structures with the network agents in $[n]$; in particular, $\hat{i}$ knows the prior $\nu(\cdot)$ as well as the signal structures $\ell_i(\cdot \mid \hat{\theta})$, $\forall \hat{\theta} \in \Theta$ and $\forall i \in [n]$, and it will make the same inference as any other agent in $[n]$ when given access to the same observations. Consider next a Gedanken experiment where $\hat{i}$ is granted direct access to all the signals of agent $i$ together with every other agents to whom agent $i$ has direct or indirect access, i.e. her neighbors and neighbors' neighbors and so on. The rate at which agent $i$ learns is then upper-bounded by the learning rate of $\hat{i}$. Formally, define $\mathcal{A}(i) = \{j \in [n] : $ there exists a path $\mathcal{P}_k(j,i)$ in $\mathcal{G}$ for some $k \in \mathbb{N}\}$. Then the Bayesian agent $i$ learns at an exponentially fast asymptotic rate that is upper bounded by

$$R_i^{\mathcal{G}}(\theta) := \min_{\check{\theta} \in \Theta \setminus \{\theta\}} D_{KL}\left(\prod_{j \in \mathcal{A}(i)} \ell_j(\cdot|\theta) \middle\| \prod_{j \in \mathcal{A}(i)} \ell_j(\cdot|\check{\theta})\right).$$

In the next section, a method of belief aggregation is proposed that applies to network topologies where each vertex has either one or no neighbors. The proposed rule is to use the same Bayesian update as in (3) if agent $i$ has no neighbors and else if agent $i$ has one neighbor, then to use (3) but with the self belief $\boldsymbol{\mu}_{i,t-1}(\cdot)$ in the right-hand side replaced by the belief $\boldsymbol{\mu}_{j,t-1}(\cdot)$ of the unique neighbor $\{j\} = \mathcal{N}(i)$. With this rule and in the case of a directed $n$-node circle where $\mathcal{A}(i) = [n]$ for any $i$, one can realize exponentially fast learning with an asymptotic rate of $(1/n)R_i^{\mathcal{G}}(\theta)$ which is within a constant factor of the above upper bound.

## IV. MEMORYLESS NETWORK UPDATES

Consider a digraph $\mathcal{G}$ satisfying $\deg(i) \in \{0,1\}, \forall i \in [n]$. The proposed rule is to use the Bayesian update in (3) if $\deg(i) = 0$, and else to use

$$\boldsymbol{\mu}_{i,t}(\hat{\theta}) = \frac{\boldsymbol{\mu}_{j,t-1}(\hat{\theta})\ell_i(\mathbf{s}_{i,t} \mid \hat{\theta})}{\sum_{\tilde{\theta} \in \Theta}\boldsymbol{\mu}_{j,t-1}(\tilde{\theta})\ell_i(\mathbf{s}_{i,t} \mid \tilde{\theta})}, \forall \hat{\theta} \in \Theta, \quad (9)$$

where $j \in [n]$ is the unique vertex $j \in \mathcal{N}(i)$. We begin the analysis of asymptotic learning with the proposed rules by the special cases of directed circle and rooted trees in Subsections IV-A and IV-B, respectively; followed by the discussion of the class of all networks with node degrees zero and one in Subsection IV-C. These updates are a special case of the *Learning without Recall* rules that we develop in a companion paper, and they can describe the behavior of Rational but Memoryless agents who share a common prior $\nu(\cdot)$ and always interpret their current and observed beliefs as having stemmed from this common prior.

### A. Directed Circles

In this subsection, we show that the update rules in (9) are particularly amiable to a circular structure, effectively achieving the upper bound derived in Subsection III-B, except for a constant multiplicative factor.

Consider a directed circle on $n$ nodes and labeled by $[n]$, in such a way that the ordered sequence $(1, 2, \ldots, n)$ constitutes a path. Fix $i \in [n]$ arbitrarily. Starting from agent $i$ at time $t$, and successively applying (9) at times $t, t-1, \ldots$ upto $t-n+1$, yields for all $t \geqslant n$ and all $j \in \{i, i-1, \ldots, 1, 0, -1, \ldots, i+1-n\}$ that:

$$\boldsymbol{\mu}_{j,t-i+j}(\theta) = \frac{l_j(\mathbf{s}_{j,t-i+j} \mid \theta)\boldsymbol{\mu}_{j-1,t-i+j-1}(\theta)}{\sum_{\tilde{\theta} \in \Theta} l_j(\mathbf{s}_{j,t-i+j} \mid \tilde{\theta})\boldsymbol{\mu}_{j-1,t-i+j-1}(\tilde{\theta})}, \quad (10)$$

when $j \geqslant 1$, and

$$\boldsymbol{\mu}_{n+j,t-i+j}(\theta) = \frac{l_{n+j}(\mathbf{s}_{n+j,t-i+j} \mid \theta)\boldsymbol{\mu}_{n+j-1,t-i+j-1}(\theta)}{\sum_{\tilde{\theta} \in \Theta} l_{n+j}(\mathbf{s}_{n+j,t-i+j} \mid \tilde{\theta})\boldsymbol{\mu}_{n+j-1,t-i+j-1}(\tilde{\theta})},$$

when $i+1-n \leqslant j < 1$. Next keep the term $\boldsymbol{\mu}_{i,t}(\theta)$ on the right-hand side of (10) and replace for each of the terms $\boldsymbol{\mu}_{j-1,t-i+j-1}(\theta)$ on its left-hand side from the successive relation at time $t-i+j-1$ until you retrieve $\boldsymbol{\mu}_{i,t-n}$ on the left hand side for time $t-n+1$. Executing this procedure leads to the following iteration which involves only the beliefs of agent $i$ at the two points in time $t$ and $t-n$.

$$\boldsymbol{\mu}_{i,t}(\theta) = \frac{\prod_{j=0}^{i-1} l_{i-j}(\mathbf{s}_{i-j,t-j} \mid \theta) \prod_{j=i}^{n-1} l_{j+1}(\mathbf{s}_{j+1,t-n+1+i-j} \mid \theta) \boldsymbol{\mu}_{i,t-n}(\theta)}{\sum_{\tilde{\theta} \in \Theta} \prod_{j=0}^{i-1} l_{i-j}(\mathbf{s}_{i-j,t-j} \mid \tilde{\theta}) \prod_{j=i}^{n-1} l_{j+1}(\mathbf{s}_{j+1,t-n+1+i-j} \mid \tilde{\theta}) \boldsymbol{\mu}_{i,t-n}(\tilde{\theta})}, \forall t \geqslant n. \quad (11)$$

Next note that starting from $\boldsymbol{\mu}_{i,0}(\theta)$ given by (7) the above is exactly the Bayesian update of agent $i$'s belief from time $t-n$ to time $t$, given that at time $t$ agent $i$ has observed the signal $\mathbf{s}_{j,t}$, at time $t-1$ agent $i-1$ has observed the signal $\mathbf{s}_{i-1,t-1}$, and so on upto the observation of signal $\mathbf{s}_{1,t-i+1}$ at time $t-i+1$, and then signal $\mathbf{s}_{n,t-i}$ followed by $\mathbf{s}_{n-1,t-i-1}$ and so on until $\mathbf{s}_{i+1,t-n+1}$ at time $t-n+1$. Hence, if we let $\hat{t} = (1/n)t$ for all $t$ belonging to the integer multiples of $n$, then Theorem 1, together with (8), implies that $\boldsymbol{\mu}_{i,\hat{t}}(\theta) \to 1$, $\mathbb{P}$-almost surely as $\hat{t} \to \infty$, at an exponentially fast asymptotic rate of

$$R^{\text{circle}}(\theta) := \min_{\check{\theta} \in \Theta \setminus \{\theta\}} D_{KL}\left(\prod_{j=1}^{n} \ell_j(\cdot|\theta) \middle\| \prod_{j=1}^{n} \ell_j(\cdot|\check{\theta})\right),$$

or equivalently that, $\boldsymbol{\mu}_{i,t}(\theta) \to 1$, $\mathbb{P}$-almost surely as $t \to \infty$, at an exponentially fast asymptotic rate of $(1/n)R^{\text{circle}}(\theta)$. Indeed, except for a penalty of constant factor $1/n$ which decreases proportionally to the network size, with the update rules in (9) one can achieve exponentially fast learning at the upper bound rate $R^{\mathcal{G}}(\theta) = R^{\text{circle}}(\theta)$.

**Remark 2.** *In the special case that all agents receive independent and identically distributed signals we have $\ell_i(\cdot|\theta) \equiv \ell_j(\cdot|\theta)$, $\forall j \in [n] \setminus \{i\}$, and therefore*

$$(1/n)R^{\text{circle}}(\theta) = (1/n)\min_{\check{\theta} \in \Theta \setminus \{\theta\}} D_{KL}\left(\ell_i^n(\cdot|\theta) \middle\| \ell_i^n(\cdot|\check{\theta})\right)$$
$$= \min_{\check{\theta} \in \Theta \setminus \{\theta\}} D_{KL}\left(\ell_i(\cdot|\theta) \middle\| \ell_i(\cdot|\check{\theta})\right) = R_i(\theta).$$

*In other words, every agent learns the true state at the same asymptotic rate as when she relies only on her own private signals. Thereby communications with the neighboring agents offer no advantages in this case. On the other hand, if the agents learn the true parameter at different private rates, $R_i(\theta)$, $i \in [n]$, then the asymptotic rate $(1/n)R^{\text{circle}}(\theta)$ would be slower than $\max_i R_i(\theta)$ but faster than $\min_i R_i(\theta)$. That is the faster agents will be slowed down by the circular communications, while the slower agents will be sped up. However, the true advantage of communications in a directed circle is apparent when some or none of the agents can lean the true parameter on their own, that is we have $\mathcal{O}_i(\theta) \neq \varnothing$ for some or all $i \in [n]$. There following the communication rules prescribed by (9), all agents would learn the true parameter exponentially fast and at a common asymptotic rate of $(1/n)R^{\text{circle}}(\theta)$, provided that $\cap_{i \in [n]} \mathcal{O}_i(\theta) = \varnothing$, or equivalently that $R^{\text{circle}}(\theta) > 0$. This way, by communicating in a circle the agent are able to benefit from each other's observations and all learn the true state of the world asymptotically, even if none are able to learn the true state on their own.*

We now shift attention to the case of networks with rooted tree topologies.

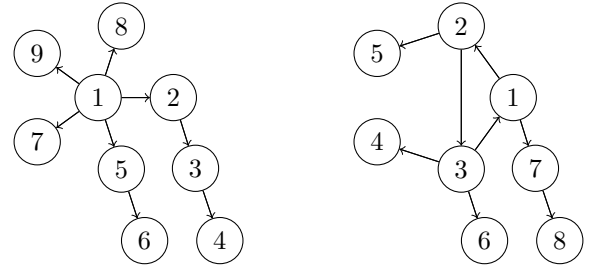

(a) A directed rooted tree  (b) A hybrid structure

Fig. 1: Some of the graph structures considered in the paper.

### B. Rooted Trees

In a directed rooted tree, a node is designated as the root and all the edges are directed away from it. Let $\mathcal{G}$ be one such directed rooted tree and label its vertices by $[n]$, assigning 1 to its root, as in example Fig. 1a. Note by the tree property that for any vertex there is a unique path connecting the root to that vertex. Take one such vertex and suppose that it is connected to the root node by a directed path consisting of $i$ distinct nodes for some $i \in [n]$. Without any loss in generality and for the ease of notation, suppose that vertices of $\mathcal{G}$ are labeled in such a way that the unique path connecting node 1 (the root) to node $i$ is given by the ordered sequence of vertices $(1, 2, 3, \ldots, i)$, as is the case for $i = 4$ in Fig. 1a.

Successive applications of (9) at times $t, t-1, t-2, \ldots, t-i+2$ for the nodes $i, i-1, i-2, \ldots, 2$ in their respective order of appearance shows that (10) applies here as well for any $j \in [i] \setminus \{1\}$. Similar to the way (11) was derived, by starting from the equation for $\boldsymbol{\mu}_{i,t}(\theta)$ and successively substituting for each of the terms $\boldsymbol{\mu}_{j-1,t-i+j-1}(\theta)$ on the left hand side, we can express the beliefs of node $i$ at each time $t \geqslant i$ in terms of the root's belief $\boldsymbol{\mu}_{1,t-i+1}$ as follows.

$$\boldsymbol{\mu}_{i,t}(\theta) = \frac{\prod_{j=2}^{i} l_j(\mathbf{s}_{j,t-i+j} \mid \theta)\boldsymbol{\mu}_{1,t-i+1}(\theta)}{\sum_{\tilde{\theta} \in \Theta} \prod_{j=2}^{i} l_j(\mathbf{s}_{j,t-i+j} \mid \tilde{\theta})\boldsymbol{\mu}_{1,t-i+1}(\tilde{\theta})}. \quad (12)$$

It is now immediate from (12) that having $\lim_{t \to \infty} \boldsymbol{\mu}_{1,t}(\theta) = 1$, $\mathbb{P}$-almost surely, is sufficient to get $\boldsymbol{\mu}_{i,t}(\theta) \to 1$ as $t \to \infty$ with $\mathbb{P}$-probability one and at the same asymptotic rate as $\boldsymbol{\mu}_{1,t}(\theta) \to 1$. In particular, we have that if the root node learns the true parameter so that per Theorem 1 we have $D_{KL}\left(\ell_1(\cdot|\theta) \| \ell_1(\cdot|\check{\theta})\right) > 0$, then for every agent $i \in [n]$ in a directed rooted tree we have that $\boldsymbol{\mu}_{i,t}(\theta) \to 1$, $\mathbb{P}$-almost surely as $t \to \infty$, at an exponentially fast asymptotic rate of $D_{KL}\left(\ell_1(\cdot|\theta) \| \ell_1(\cdot|\check{\theta})\right)$. This in the main part is due to the fact that a point mass is a stationary point for the belief iterations proposed in (9). In what follows, we shall combine the results from this and the previous subsection to address the general class of digraphs with zero or one node degrees.

## C. Generalization to Hybrid Structures

We begin by the observation that any weakly connected digraph $\mathcal{G}$ which has only degree zero or degree one nodes can be drawn as a rooted tree whose root is replaced by a directed circle, e.g. Fig. 1b. This is true since any such digraph can have at most one directed circle and all other nodes that are connected to this circle should be directed away from it, otherwise $\mathcal{G}$ would have to include a node of degree two or higher.

The case of digraphs $\mathcal{G}$ with no circles is the rooted trees discussed in Subsection IV-B. Therefore suppose that $\mathcal{G}$ contains a circle of length $l$ consisting of its first $l$ nodes and let them be labeled by $[l]$. Next note that any of the $l$ nodes belonging to the directed circle would learn the true state of the world at the exponentially fast asymptotic rate of $(1/l)R^{\text{circle}}(\theta)$ derived in Subsection IV-A. Moreover, each of the nodes $i \in [n]\setminus[l]$ is connected uniquely to a node $i_l \in [l]$ and along a distinct path $(i_l, i_{l_1}, i_{l_2}, \ldots, i_{l_k}, i)$ for some $k \in [n-l-1]$. Thereby, using the same argument as the one leading to (12) in the case of rooted trees, we get that for any $i \in [n]\setminus[l]$ that

$$\boldsymbol{\mu}_{i,t}(\theta) = \frac{l_i(\mathbf{s}_{i,t} \mid \theta) \prod_{j=1}^{k} l_{i_j}(\mathbf{s}_{i_j, t-k+j-1} \mid \theta) \boldsymbol{\mu}_{i_l, t-k-1}(\theta)}{\sum_{\tilde{\theta} \in \Theta} l_i(\mathbf{s}_{i,t} \mid \tilde{\theta}) \prod_{j=1}^{k} l_{i_j}(\mathbf{s}_{i_j, t-k+j-1} \mid \tilde{\theta}) \boldsymbol{\mu}_{i_l, t-k+1}(\tilde{\theta})},$$

wherefore with $\mathbb{P}$-probability one as $t \to \infty$ if $\boldsymbol{\mu}_{i_l,t}(\theta) \to 1$, then $\boldsymbol{\mu}_{i,t}(\theta) \to 1$ as well, and at the same asymptotic rate. Indeed, we have that if $R^{\text{circle}}(\theta) > 0$, then every agent in the network would learn the true state $\theta$ asymptoticly exponentially fast, and all at the same rate given by $(1/l)R^{\text{circle}}(\theta) > 0$.

*A Leader-Follower Architecture:* The preceding results can be summarized upon the observation that those agent belonging to the so-called "root circle" combine their observations in a Bayesian manner (except for a penalty of $1/l$ in the asymptotic rate) and once the opinions of the circle agents converge to a point mass, the rest of the agents follow as well, after a finite number of steps that depends on their distance to the root circle. Indeed, the first $l$ agents form a circle of leaders where they combine their observations and reach a consensus; every other agent in the network then follows whatever state that the leaders have collectively agreed upon.

## V. Concluding Remarks

In this paper, a belief aggregation method is proposed and shown to be applicable to a class of directed networks that can be drawn as a rooted tree with the root node replaced by a directed circle. The proposed update rules replicate that of a single Bayesian agent except that in the case of degree one nodes the self-beliefs are replaced by the beliefs communicated by the neighboring agents. Accordingly, those agents which belong to the root circle can combine their observations to learn the true state of the world, even if none can distinguish the truth privately. Any peripheral agent that does not belong to the root circle would then follow the beliefs of the root agent to whom she is connected either directly or indirectly through her neighbor and neighbor's neighbor etc. Thereby, all agents in the network would learn the true state of the world exponentially fast and at the same asymptotic rate, so long as the truth is distinguishable through the combined observations of all agents in the root circle. The asymptotic rate at which learning occurs is shown to equal $(1/l)R^{\text{circle}}$, where $l$ is the length of the root circle, and $R^{\text{circle}}$ is the exponential rate at which a Bayesian agent with direct access to all the observations of the root agents would learn the truth. The authors' ongoing research focuses on the investigation and analysis of belief update rules that provide asymptotic learning in a wider variety of network structures and facilitate the tractable modeling and analysis of rational behaviors in networks, using the so-called *Learning without Recall* framework.